\documentclass[12pt]{article}

\def\rdots{\mathinner{\mkern1mu\raise1pt\vbox{\kern1pt\hbox{.}}\mkern2mu
   \raise4pt\hbox{.}\mkern2mu\raise7pt\hbox{.}\mkern1mu}}
\newcommand{\Z}{{\rm Z\kern-.35em Z}}
\newcommand{\bP}{{\rm I\kern-.15em P}}
\newcommand{\Q}{\kern.3em\rule{.07em}{.65em}\kern-.3em{\rm Q}}
\newcommand{\R}{{\rm I\kern-.15em R}}
\newcommand{\h}{{\rm I\kern-.15em H}}
\newcommand{\C}{\kern.3em\rule{.07em}{.55em}\kern-.3em{\rm C}}
\newcommand{\T}{{\rm T\kern-.35em T}}
\newcommand{\D}{{\kern-.5em /}}

\begin{document}

\openup 1.5\jot

\centerline{Chiral Anomalies? \ \ A Voice for Choice $^*$}

\vspace{1in}
\centerline{Paul Federbush}
\centerline{Department of Mathematics}
\centerline{University of Michigan}
\centerline{Ann Arbor, MI 48109-1003}
\centerline{(pfed@math.lsa.umich.edu)}
\vspace{4.0in}

$^*$ This work was supported in part by the National Science Foundation under Grant No. PHY-92-04824 .
\vfill\eject

$$ $$
$$ $$

\centerline{{\bf ABSTRACT}}

\vspace{1in}

\indent

We consider theories with  gauged chiral fermions in which there are abelian anomalies, and no nonabelian anomalies (but there may be nonabelian gauge fields present).  We construct an associated theory that is gauge invariant, renormalizable, and with the same particle content, by adding a finite number of terms to the action.  Alternatively one can view the new theory as arising from the original theory by using another regularization, one that is gauge invariant.  The situation is reminiscent of the mechanism of adding Fadeev-Popov ghosts to an unsatisfactory gauge theory, to arrive at the usual quantization procedure.  The models developed herein are much like the abelian Wess-Zumino model (an abelian effective theory with a Wess-Zumino counter term), but unlike the W-Z model are renormalizable!  

\vfill\eject

In two recent papers I have presented a gauge-invariant regularization of the Weyl determinant using wavelets [1].  In that work cutoff gauge-invariant expressions were obtained, but the task of letting the cutoff go to infinity was left open.  The expressions were also unwieldy and not computationally effective.  The present paper assumes no knowledge of this previous work!  The wavelet constructions may be viewed as only motivating the search for gauge-invariant theories such as here presented.

The theory of the current paper is rather similar to the abelian Wess-Zumino model (the abelian effective gauge theory with a Wess-Zumino counter term) [2], though the development is along slightly different lines.  It may be viewed as a ``minimal" modification of this Wess-Zumino model making that theory renormalizable.  One arrives at a new mechanism to construct theories that were once rejected as possessing anomalies.  We expect many applications of the present construction.   (Nothing here, of course, prevents the $\pi^0$ from decaying into two photons, the $\pi^0$ field given by the same operator as always.)

We first consider the computation of the Weyl determinant for a chiral fermion in a classical (unquantized) gauge field.  The gauge group is arbitrary, may be abelian or nonabelian, irreducible or reducible.  The representation is also arbitrary, need not satisfy the no-anomaly condition!  But we assume only abelian anomalies.    We impose the following conditions on the gauge fields:

\bigskip
\begin{description}
\item[1)]  The $A_\mu(x)$ are ``small''.  So that we are in the perturbative regime.
\item[2)]  The $A_\mu(x)$ are zero at infinity.   So that there are no infrared difficulties.
\end{description}

\bigskip

We will construct an expression for the Weyl determinant $\det W$ that satisfies the following three properties:

\bigskip
\begin{description}
\item[P1)]  The determinant is Lorentz invariant.  (Alternatively, one could perform a Euclidean construction.)
\item[P2)]  The determinant is gauge-invariant (under gauge transformations that are the identity near infinity).
\item[P3)]  The expansion for $\ell n(\det W)$ differs from the ``usual expansion'' by ``added terms'' that all vanish for the gauge field in the Lorentz gauge.
\item[P4)] The expansion for $\ell n(\det W)$ is a ``regularization'' of the usual perturbation expansion.  (This is clarified below.)
\item[P5)] The ``added terms'' occur only for degree $n \le 4$, i.e. associated to the ``divergent'' triangle and box diagrams.
\end{description}
$$ $$

Properties relating to quantization of the gauge field will be discussed later.

We consider the expansion of $\ell n(\det W)$ in powers of the field (as mentioned above we work in the perturbative regime)
\begin{equation}
\ell n(\det W) = T_1 + T_2 + \cdots
\end{equation}
\begin{equation}
T_n = \int d^4x_1 \cdots \int d^4x_n \ f^{\mu_1 \cdots \mu_n}_{n \; a_1 \cdots a_n} (x_1, \cdots, x_n) A^{a_1}_{\mu_1} (x_1) \cdots A^{a_n}_{\mu_n}(x_n).
\end{equation}
Summation over group indices, $a_i$, and vector indices, $\mu_i$, is assumed.  The $f_n$ are actually distributions.  Perturbation theory yields finite expressions for $f_n$ if $n > 4$.  Perturbation theory yields finite expressions for all the $f_n$ for values of the $x_i$ that are distinct (non-coincident arguments).  A ``regularization'' of $T_n$ is a choice of an $f_n$, defined for all values of its arguments, that agrees with the perturbation theory result for distinct arguments.  This is the general definition of a ``regularization'', but many physicists work with more restrictive definitions (using only certain operations on the Feynman integrals invovled).  We use the general definition above.  Our ``regularization'' will differ from the ``usual regularization'' only for $n \le 4$.

We abbreviate (2) as
\begin{equation}
T_n = \int dx_1 \cdots \int dx_n \; f_n \; A(x_1) \cdots A(x_n).
\end{equation}
We use $T_n, f_n$ to denote ``our theory'' and $\hat{T}_n, \; \hat{f}_n$ to denote the ``usual theory''.  These will differ by the ``added terms'' we have put in our Action.  We also write
\begin{eqnarray}
f_n &=& \hat{f}_n + d_n \\
T_n &=& \hat{T}_n + D_n.
\end{eqnarray}
The $d_n$ or $D_n$ are the ``added terms''.  So $d_n = 0$ for $n > 4$.

Since $f_n$ and $\hat{f}_n$ are both ``regularizations'' of the same perturbation expression, $d_n$ will be zero if all its arguments are distinct (there must be coincident arguments for $d_n$ to be nonzero).

The ``usual regularization'' for the singlet chiral anomaly situation is as derived by Adler, Bell, and Jackiw in [3].  (The ``usual regularization'' in general is as given by Bardeen [4].)  We will define our theory by finding the correct added terms, $d_n$, to add to the usual regularizations.  (The work of [1] shows that it is possible to arrive at such gauge-invariant theories directly, rather than by ``correcting'' the usual approach, as is the route here followed.)

We first define our gauge-invariant regularization for the abelian gauge group case.  We need only deal with the famous AVV and AAA triangle graphs first treated in [3], and now analyzed in many quantum field theory textbooks.  It is sufficient to explain the AVV terms.  The usual regularization of the AVV diagram contributes to $\hat{f}_3$ a term
\begin{equation}
M_{\alpha\beta\gamma}(x, \; y, \; z)
\end{equation}
with
\begin{equation}
{\partial \over \partial y_\beta} \; M = {\partial \over \partial z_\gamma} \; M = 0.
\end{equation}
We define a corresponding ``added term'' in $d_3$ to be
\begin{equation}
c \; {\partial \over \partial x_\alpha} \; D(x,y)\; \delta(y-z) \; \varepsilon_{\beta\gamma ij} \ {\partial \over \partial y_i} \ {\partial \over \partial z_j} .
\end{equation}
$D$ is as usual ${1 \over k^2}$ in momentum space.  With correct choice of $c$ the divergence of $f_3$ will be zero at all three vertices.  Note that $d_3$ has contributions only for coincident arguments (it defines a difference between two regularizations), and that the corresponding $D_3$ will vanish if the fields are in the Lorentz gauge (integrating by parts the ${\partial \over \partial x_\alpha}$ derivative) as stated in P3).

We view the $D$ propagator in (8) as associated to a ``ghost particle'', somewhat analogous to the Fadeev-Popov ghosts.  With quantization of the gauge field we would desire that these ghost particles do not contribute in intermediate states of the unitarity relationship (as with the F-P ghosts).

We now state the sixth property we assert.

\begin{description}
\item[P6)] The ``ghost particles'' defined present in the ``added terms'' do not contribute as particles in intermediate states of the unitarity relationship.
\end{description}

The derivative attached to $D$ (as a ghost intermediate state particle) always hits a conserved vertex, in a loop or a line, and thus such terms make no contribution.  (It is technically convenient to let the gauge meson have a small mass during this demonstration.)  This argument is of the same type as standard in studying the unitarity relationship for gauge theories.

The usual singlet chiral anomaly may be similarly analyzed to the preceding discussion (there will be ``added terms'' both in $d_3$ and $d_4$).

The abelian W-Z model studied by Preskill in [2] has added terms in the Lagrangian 
\begin{equation}
{\rm Tr}(F\tilde{F})\phi \ + \ c(\partial_\mu \phi - A_\mu)^2
\end{equation}
with $\phi$ an introduced scalar field.  Our added term in the action (whose kernel is given in (8) may be written as
\begin{equation}
c \int dx \; dy {\rm Tr} \left( F(x) \tilde{F}(x) \right) D(x,y) \partial_\mu A_\mu(y)
\end{equation}
It is easily seen that integrating out the field $\phi$ in (9) gives rise to the terms (10), as well as other terms that lead to the non-renormalizability of this W-Z model.  We are keeping a renormalizable part of the W-Z model.

Contrary to the claims of an earlier version of this paper, the situation in the case of a nonabelian anomaly remains unclear.  We do not know if a similar mechanism to that employed herein on abelian anomalies can be extended to nonabelian anomalies, but we remain hopeful it can, and are working on this problem.

$$ $$

\noindent
ACKNOWLEDGEMENT.  I would like to thank H. Greorgi for a communication concerning the relationship to Wess-Zumino models.

\vfill\eject

\centerline{{\bf REFERENCES}}

[1]

\noindent
P. Federbush, ``A New Formulation and Regularization of Gauge Theories Using a Non-Linear Wavelet Expansion'', to be published in {\it Prog. Theor. Phys.}, Vol. {\bf 94} No. 6 (1995) (on-line hep-ph/9505368).

\noindent
P. Federbush, ``A Gauge-Invariant Regularization of the Weyl Determinant Using Wavelets'', preprint (on-line hep-th/9509131).

[2]

\noindent
J. Wess and B. Zumino, ``Consequences of Anomalous Ward Identities", {\it Phys. Lett. } {\bf B37}, 95 (1971).

\noindent
J. Preskill, ``Gauge Anomalies in an Effective Field Theory", {\it Ann. Phys.} {\bf 210}, 323 (1991).

[3]

\noindent
S.L. Adler, ``Axial-Vector Vertex in Spinor Electrodynamics'', {\it Phys. Rev.} {\bf 177}, 2426 (1969).

\noindent
J.S. Bell and R. Jackiw, `` A PCAC Puzzle: $\pi^0 \rightarrow \gamma \gamma$ in the Sigma Model'', Nuovo Cim. {\bf 60A}, 47 (1967).

\noindent
S.L. Adler, ``Perturbation Theory Anomalies'', in \underline{Lectures in Elementary Particle Physics}, ed. S. Deser, M. Grisaru and H. Pendleton (M.I.T. Press, Cambridge, MA.) (1970).

[4]

\noindent
W.A. Bardeen, ``Anomalous Ward Identities in Spinor Field Theories'', {\it Phys. Rev.} {\bf 184}, 1848 (1969).

\end{document}